# Quantum theory of Rayleigh scattering

A. P. Vinogradov,[1,2] V. Yu. Shishkov,[1,2] I. V. Doronin,[1,2] E. S. Andrianov,[1,2] A. A. Pukhov,[1,2] and A. A. Lisyansky[3,4]

[1]*Dukhov Research Institute of Automatics (VNIIA), 22 Sushchevskaya, Moscow 127055, Russia*
[2]*Institute for Theoretical and Applied Electromagnetics, 13 Izhorskaya, Moscow 125412, Russia*
[3]*Department of Physics, Queens College of the City University of New York, Flushing, New York 11367, U.S.A.*
[4]*The Graduate Center of the City University of New York, New York, New York 10016, U.S.A.*

**Abstract:** We develop a quantum theory of atomic Rayleigh scattering. Scattering is considered as a relaxation of incident photons from a selected mode of free space to the reservoir of the other free space modes. Additional excitations of the reservoir states which appear are treated as scattered light. We show that an entangled state of the excited atom and the incident photon is formed during the scattering. Due to entanglement, a photon is never completely absorbed by the atom. We show that even if the selected mode frequency is incommensurable with any atomic transition frequency, the scattered light spectrum has a maximum at the frequency of the selected mode. The linewidth of scattered light is much smaller than that of the spontaneous emission of a single atom, therefore, the process can be considered as elastic. The developed theory does not use the phenomenological concept of "virtual level."

## 1. Introduction

The internal structure of various materials is probed with a beam of particles (electrons, neutrons, ions, and photons). Photons are most commonly used to determine the structure of atoms and molecules.

The history of this method dates back to 1869 when Tyndall [1] observed light scattered by very small particles. In 1915, Cabannes [2] observed elastic scattering from gasses. The theory for Tyndall's observations was developed by Rayleigh [3], who treated the problem of elastic scattering of an electromagnetic (EM) wave by a subwavelength particle as forced oscillations of the polarization of the subwavelength particle. The theory predicted symmetry in forward and backward scattering of light from a single particle, the polarization of the scattered light, and successfully explained the blue color of the sky.

However, experimental data on the observation of bright lines in the solar spectrum[4, 5] were not explained by this theory. Such lines indicated that the phenomenon is of resonant nature. Since the atoms are of subwavelength sizes, classical Rayleigh's theory predicts no resonant phenomena. The nature of these lines was only understood with the development of quantum mechanics [5], in which the lines were associated with the resonant transitions between the eigenstates of the electronic subsystems of atoms or molecules accompanied by emission of photons. The quantum-mechanical study of the eigenstates of atoms and transitions between them resulted in the development of optical spectroscopy. In contrast to Rayleigh's theory, the optical spectroscopy deals with resonance phenomena.

Since optical spectroscopy deals with resonances, its adaptation to the scattering problem inevitably has required phenomenological assumptions. First of all, since the light scattering was treated as a sequence of photon absorption and emission [6, 7], the theory introduced an additional virtual state with a transition frequency which coincides with the frequency of scattered light. Although such an assumption gives a visual picture of the phenomenon, there is no evidences to support the introduction of the virtual level. But via this *ad hoc* assumption the scattering problem was reduced to the problem of absorption and emission of photons, in which the quantum system transitions between eigenstates. This assumption contains internal contradictions.

Since the linewidth of Rayleigh scattering coincides with the linewidth of the incident light and the absorption-emission process includes spontaneous emission from the virtual state, the latter should have a very narrow linewidth or a very long lifetime. But the experiments show that the lifetime of the process is very short.

Following the pioneering work of Placzek [8], the Kramers-Heisenberg formula [9] is used to calculate the scattering cross section (see also [10] and [11]). This formula corresponds to the second-order perturbation in the interaction between an incident photon with an atom and implies that the initial atomic state coincides with its final state, while the state of the photon (e.g., the direction of the emitted photon wavevector or the polarization) may be changed. The need for such an approach was dictated by the fact that the first order perturbation theory gives a result of zero. In addition, it was assumed that the reverse process also contributes to the scattering cross section - first, a photon is emitted by an excited atom, and then the atom, which is already unexcited, absorbs a photon [6].

Such an approach raises some questions. Placzek's theory implies that during the scattering process, the incident photon and the atom are in a superposition state, excited level of which is populated with some probability amplitude. Thus, there should be a superposition state at which the incident photon is absorbed. In this state, the dipole moment of the atom is nonzero, and the atom should radiate. The absorption-emission theory *postulates* [6] that since the scattering is elastic emitted photon *must* have the same frequency as the absorbed one. To substantiate this postulate and simplify the treatment, the problem was reduced to the problem of resonant scattering [6]. In this case, elastic scattering follows from Fermi's golden rule. The generalization to the case of a photon with frequency incommensurable with atomic transitions requires the introduction of a virtual level with a transition frequency equal to the frequency of the incident light. Thus, the superposition state is replaced by a state with an excited virtual level. In particular, a photon can be captured to the virtual level, with the subsequent emission of a photon, so that the system returns to the ground state. In this approach, the scattering becomes a second-order radiation process. In this paper, we show that if instead of a virtual state of the atom, one consistently considers a superposition state of an incident photon and the atom, the scattering problem yields correct results in the first order of the perturbation theory.

In the absorption-emission theory, however, the magnitude of the transition dipole element and the lifetime of the virtual level are unknown. Another problem, which this theory does not solve, is the coherence of Rayleigh scattering. The spontaneous emission of atoms is incoherent and has a finite linewidth. To explain their coherence, the presence of a fixed phase in a quantum state is *postulated* [6]. This phase should not depend on the preceding evolution of the state. In this case, all atoms radiate in-phase, and the radiation intensity is quadratic in the number of atoms. This not only implies the existence of an operator of the phase, but also that this operator must commute with the Hamiltonian of the atom. As far as we know, no such operator exists.

Although the quantum theory of the Rayleigh scattering has almost a century-long history, it still contains unclear points and contradictions. To clear up the situation it is necessary to introduce new ideas.

In this context, it is worth noting that the further development of quantum optics made it possible to eliminate the above-mentioned contradictions in Placzek's theory and to avoid unfounded assumptions. This progress is based on the following works.

First, in 1937, Rabi [4] considered the precession of a magnetic dipole moment in a magnetic field and showed that the probability of changing the spin direction of an atom to the opposite oscillates. Thus, for a quantum system, there appears an alternative for the evolution under the influence of an external force. Not only can it transition between the eigenstates but also be driven by an external force.

Second, Jaynes and Cummings [12] considered the interaction of an atom with a single mode of an EM field found eigenstates of this system. This implies that the transitions of the whole system between the Jaynes–Cummings states are responsible for radiation and not the transitions between the atom eigenstates. The system may radiate and transition between its eigenstates, only if the third "participant" is needed, which may cause the transitions. This may be the reservoir of free space modes or a third body like an electron or another atom. Without this participant, the interaction of an atom with a single mode reduces to Rabi oscillations [13]. Thus, a radiating atom is a system which interacts with a radiative reservoir, i.e., the atom becomes an open quantum system. Therefore, the third important moment is the introduction by Lax [14] of the Heisenberg-Langevin equations for the evolution of quantum operators and the master equation for the evolution of the density matrix [15, 16].

Finally, the invention of the laser significantly changed both the experiment and its theoretical description. High intensity laser light (a large number of photons) makes possible the use of a classical description of an incident EM wave because, during a single act of absorption or emission of a photon, the total number of photons changes insignificantly. The coherent properties of laser light are close to these of a classical EM wave. Thus, laser light may be treated as a classical field with constant amplitude.

All these ideas permit to go beyond Fermi's golden rule and describe the system as an open quantum system applying the master or Heisenberg-Langevin equations.

In this paper, we develop a quantum theory of Rayleigh scattering of light by an atom. The eigenstates of the system include both the atom and incident photons from a selected mode of the EM field of free space. In such eigenstates, the atomic state and state of the photon belonging to the selected mode are entangled. Therefore, a photon is never completely absorbed by the atom. This theory does not require an introduction of an additional artificial virtual level. In the developed theory, during the scattering, an atomic level in excited nonresonantly. Consequently, in a superposition state, the atomic dipole moment is nonzero, and the process of re-emission of the photon takes place from the very beginning. We show that the probability amplitudes of the excited atomic state oscillate with the frequency of the incident photon. The excitations appearing in the reservoir due to the incident photon interaction with the free space modes are treated as scattering photons. The spectrum of these photons has a maximum at the frequency of the selected mode, and the linewidth of the maximum is much narrower than the linewidth of the spontaneous emission from a single atom. The results of the absorption-emission theory [6] are critically discussed.

## 2. Rayleigh scattering as relaxation of the system "selected mode + atom" into free space modes

In further consideration, we assume that the frequency of the incident photon is incommensurable with any frequency of inter-level atomic transitions. This allows us to represent atoms as two-level systems (TLSs).

Consider an atom placed in free space. Following the traditional scheme of quantization of the EM field, we assume that the volume $V$ of the space in which quantization occurs is large but finite. We represent the atom by a TLS with the ground state $|g\rangle$, the excited state $|e\rangle$, and the transition frequency $\omega_{TLS}$. We assume that the EM modes of this volume are in thermal equilibrium with a given temperature $T$. The eigenstates of the modes are described by Fock states $|n\rangle$ with a given number of photons in the mode.

The object of our investigation is a quantum atom interacting with a selected free space mode, to which $n_0$ incident photons belong at the initial moment of time. The rest of the free space modes are considered as a reservoir. The number of photons $n_0$ in this *selected mode* determines the energy of the incident wave as $\hbar\omega_{SM} n_0$. Other EM field modes are assumed to be empty. The scattering process is considered as the relaxation of photons in the selected mode into the reservoir. Additional excitations of the reservoir modes, which appear during the relaxation, are treated as scattered light.

We solve the scattering problem in two steps. First, we focus on the subsystem consisting of the atom and a selected mode with the wave vector $\mathbf{k}_{SM}$, the polarization $\lambda_{SM}$, and the frequency $\omega_{SM} = c|\mathbf{k}_{SM}|$. Suppose that this selected mode is excited, and there are $n_0$ photons in it. Eigenfrequencies of this mode, $\omega_{SM}(n+1/2)$, $n = 0,1,2,...$, form an equidistant spectrum. Thus, initially, the selected mode is not in thermal equilibrium with the other free space modes. This mode plays the role of incident radiation. At this stage, we neglect the interaction of the atom with empty modes.

In the absence of the interaction between the modes and the atom, the system eigenstates are direct products of the eigenstates of the modes and the atom, $|n,e\rangle$ and $|n,g\rangle$. When the interaction is taken into account, the atom and photon states are entangled. Following Jaynes–Cummings, we refer to them as $|+,n\rangle$ and $|+,n\rangle$ (see for detail [12]).

The interaction between the atom and the mode is considered in the dipole and rotating wave approximations. Then, the Hamiltonian of this subsystem has the form of the James-Cummings Hamiltonian [13, 17]:

$$\hat{H}_S = \hbar\omega_{TLS}\hat{\sigma}^\dagger\hat{\sigma} + \hbar\omega_{SM}\hat{a}^\dagger\hat{a} + \hbar\Omega_R\left(\hat{a}^\dagger\hat{\sigma} + \hat{\sigma}^\dagger\hat{a}\right), \qquad (1)$$

where $\Omega_R = -\mathbf{E}_0\cdot\mathbf{d}/\hbar = -\sqrt{2\pi\hbar\omega_{SM}/V}\mathbf{e}_{\mathbf{k}_0,\lambda_0}\cdot\mathbf{d}/\hbar$ is the interaction constant (the Rabi frequency), $\mathbf{e}_{\mathbf{k}_0,\lambda_0}$ is the unit polarization vector of the selected mode, $\mathbf{d}$ is the matrix element of the dipole transition of the TLS, $\sigma = |g\rangle\langle e|$ and $\sigma = |e\rangle\langle g|$ are lowering and raising operators for the TLS, respectively, and $\hat{a}$ and $\hat{a}^\dagger$ are photon annihilation and creation operators, respectively.

The eigenstates of Hamiltonian (1) are well-known [12] (see also [13, 18]). Among them, there is the ground state that does not contain photons ($n = \langle\hat{a}^\dagger\hat{a}\rangle = 0$)

$$|g,0\rangle. \qquad (2)$$

We assume that the energy of this state is zero. Excited eigenstates can be combined in pairs:

$$\begin{aligned}|+,n\rangle &= \cos\varphi_n|e,n-1\rangle + \sin\varphi_n|g,n\rangle,\\|-,n\rangle &= -\sin\varphi_n|e,n-1\rangle + \cos\varphi_n|g,n\rangle,\ n=1,2,\ldots;\end{aligned} \qquad (3)$$

where $\varphi_n = \tan^{-1}\left(2\Omega_R\sqrt{n}/|\Delta|\right)/2$. The eigenfrequencies of these states are

$$\omega_{\pm,n} = (n-1)\omega_{SM} + (\omega_{SM}+\omega_{TLS})/2 \pm \sqrt{\Omega_R^2 n + (\Delta/2)^2},\ n=1,2,\ldots, \qquad (4)$$

where $\Delta = \omega_{SM} - \omega_{TLS}$ is the frequency detuning. The value $2\sqrt{\Omega_R^2 n + (\Delta/2)^2}$ is the frequency difference between the states $|+,n\rangle$ and $|-,n\rangle$.

We assume that the selected mode initially contains $n_0$ photons, and the atom is in the ground state so that the system initial state is $|n_0,g\rangle$. This state is not the eigenstate of the "selected mode + atom" system. Therefore, as we show below, oscillations of nondiagonal elements of the system density matrix begin.

In the second step, we consider the scattering process. In the formalism described above, this process is reduced to the evolution of the system "selected mode + atom," which interacts with other free space modes. We emphasize that the relaxation of the system occurs due to the interaction of the atom with the free space modes, while the selected mode does not interact directly with the other free space modes. The excitation of free space modes is treated as scattering.

The Hamiltonian of the free space modes has the form [6, 13]:

$$\hat{H}_R = \sum_{\mathbf{k},\lambda}\hbar\omega_{\mathbf{k},\lambda}\hat{a}_{\mathbf{k},\lambda}^\dagger\hat{a}_{\mathbf{k},\lambda}, \qquad (5)$$

where $\omega_{\mathbf{k},\lambda}$ is the frequency of the free space mode with the wave vector $\mathbf{k}$ and the polarization $\lambda$, $\hat{a}_{\mathbf{k},\lambda}$ and $\hat{a}_{\mathbf{k},\lambda}^\dagger$ are annihilation and creation operators, respectively. The atomic dipole moment interacts with these modes. The corresponding interaction Hamiltonian has the form

$$\hat{H}_{SR} = \sum_{\mathbf{k},\lambda} \hbar \gamma_{\mathbf{k},\lambda} \left( \hat{a}^\dagger_{\mathbf{k},\lambda} \hat{\sigma} + \hat{\sigma}^\dagger \hat{a}_{\mathbf{k},\lambda} \right), \tag{6}$$

where $\gamma_{\mathbf{k},\lambda} = -|\mathbf{d}| \sqrt{2\pi \hbar \omega_{\mathbf{k},\lambda} / V} / \hbar$ is the interaction constant of the atomic dipole moment with the electric field of the free space mode. The form of Hamiltonian (6) indicates that the interactions of an atom with each mode are independent. Further, it is convenient to rewrite the operator of the interaction of the system with reservoir (6) in the basis of the eigenstates of the "atom + mode" system, i.e. in basis (2)-(3). One can show that in this basis, the operator $\hat{\sigma}$ has the form

$$\hat{\sigma} = \cos\varphi_1 \hat{S}_{(g,0)(+,1)} - \sin\varphi_1 \hat{S}_{(g,0)(-,1)} + \sum_{n=1}^{\infty} (\cos\varphi_n \sin\varphi_{n-1} \hat{S}_{(+,n-1)(+,n)} - \\ - \sin\varphi_n \sin\varphi_{n-1} \hat{S}_{(+,n-1)(-,n)} + \cos\varphi_n \cos\varphi_{n-1} \hat{S}_{(-,n-1)(+,n)} - \sin\varphi_n \cos\varphi_{n-1} \hat{S}_{(-,n-1)(-,n)}, \tag{7}$$

and the Hamiltonian $\hat{H}_{SR}$ of the interaction of the system with the reservoir can be written as

$$\hat{H}_{SR} = \sum_{\mathbf{k},\lambda} \hbar \gamma_{\mathbf{k},\lambda} \hat{a}^\dagger_{\mathbf{k},\lambda} \Big( \cos\varphi_1 |g,0\rangle\langle +,1| - \sin\varphi_1 |g,0\rangle\langle -,1| \\
+ \sum_{n=1}^{\infty} \big( \cos\varphi_n \sin\varphi_{n-1} |+,n-1\rangle\langle +,n| - \sin\varphi_n \sin\varphi_{n-1} |+,n-1\rangle\langle -,n| \\
+ \cos\varphi_n \cos\varphi_{n-1} |-,n-1\rangle\langle +,n| - \sin\varphi_n \cos\varphi_{n-1} |-,n-1\rangle\langle -,n| \big) \Big) + \text{h.c.} \\
= \sum_{\mathbf{k},\lambda} \hbar \gamma_{\mathbf{k},\lambda} \hat{a}^\dagger_{\mathbf{k},\lambda} \Big( \cos\varphi_1 \hat{S}_{(g,0)(+,1)} - \sin\varphi_1 \hat{S}_{(g,0)(-,1)} \\
+ \sum_{n=1}^{\infty} \big( \cos\varphi_n \sin\varphi_{n-1} \hat{S}_{(+,n-1)(+,n)} - \sin\varphi_n \sin\varphi_{n-1} \hat{S}_{(+,n-1)(-,n)} \\
+ \cos\varphi_n \cos\varphi_{n-1} \hat{S}_{(-,n-1)(+,n)} - \sin\varphi_n \cos\varphi_{n-1} \hat{S}_{(-,n-1)(-,n)} \big) \Big) + \text{h.c.} \tag{8}$$

where we introduce the notations for the operators of possible transitions between the eigenstates of the subsystem:

$$\hat{S}_{(+,n-1)(+,n)} = |+,n-1\rangle\langle +,n|, \hat{S}_{(+,n-1)(-,n)} = |+,n-1\rangle\langle -,n|, \\
\hat{S}_{(-,n-1)(+,n)} = |-,n-1\rangle\langle +,n|, \hat{S}_{(-,n-1)(-,n)} = |-,n-1\rangle\langle -,n|, \\
\hat{S}_{(g,0)(-,1)} = |g,0\rangle\langle -,1| \tag{9}$$

Knowing the eigenstates of the "selected mode + atom" system, it is possible to exclude free space mode variables with the exception of the selected mode. For this, it is necessary to assume that the reservoir of free space modes is in a state of thermal equilibrium at a given temperature. Neglecting relativistic effects, we can assume that the frequencies are limited to the optical range. Further, we assume that $T = 0$, i.e., the free space modes are empty. Following the standard procedure, we obtain the density matrix in basis (3), which obeys the master equation of the Lindblad form [13, 19, 20]:

$$\partial \hat{\rho} / \partial t = \frac{i}{\hbar}\left[\hat{\rho}, \hat{H}_S\right] +$$

$$+ \frac{\gamma_0}{2} \cos^2 \varphi_1 \left(2\hat{S}_{(g,0)(+,1)} \hat{\rho} \hat{S}^\dagger_{(g,0)(+,1)} - \hat{S}^\dagger_{(g,0)(+,1)} \hat{S}_{(g,0)(+,1)} \hat{\rho} - \hat{\rho} \hat{S}^\dagger_{(g,0)(+,1)} \hat{S}_{(g,0)(+,1)}\right)$$

$$+ \frac{\gamma_0}{2} \sin^2 \varphi_1 \left(2\hat{S}_{(g,0)(-,1)} \hat{\rho} \hat{S}^\dagger_{(g,0)(-,1)} - \hat{S}^\dagger_{(g,0)(-,1)} \hat{S}_{(g,0)(-,1)} \hat{\rho} - \hat{\rho} \hat{S}^\dagger_{(g,0)(-,1)} \hat{S}_{(g,0)(-,1)}\right) +$$

$$+ \frac{\gamma_0}{2} \sum_{n=1}^{\infty} \Big( \cos^2 \varphi_n \sin^2 \varphi_{n-1} \times$$

$$\times \left(2\hat{S}_{(+,n-1)(+,n)} \hat{\rho} \hat{S}^\dagger_{(+,n-1)(+,n)} - \hat{S}^\dagger_{(+,n-1)(+,n)} \hat{S}_{(+,n-1)(+,n)} \hat{\rho} - \hat{\rho} \hat{S}^\dagger_{(+,n-1)(+,n)} \hat{S}_{(+,n-1)(+,n)}\right) +$$

$$+ \sin^2 \varphi_n \sin^2 \varphi_{n-1} \left(2\hat{S}_{(+,n-1)(-,n)} \hat{\rho} \hat{S}^\dagger_{(+,n-1)(-,n)} - \hat{S}^\dagger_{(+,n-1)(-,n)} \hat{S}_{(+,n-1)(-,n)} \hat{\rho} - \hat{\rho} \hat{S}^\dagger_{(+,n-1)(-,n)} \hat{S}_{(+,n-1)(-,n)}\right)$$

$$+ \cos^2 \varphi_n \cos^2 \varphi_{n-1} \left(2\hat{S}_{(-,n-1)(+,n)} \hat{\rho} \hat{S}^\dagger_{(-,n-1)(+,n)} - \hat{S}^\dagger_{(-,n-1)(+,n)} \hat{S}_{(-,n-1)(+,n)} \hat{\rho} - \hat{\rho} \hat{S}^\dagger_{(-,n-1)(+,n)} \hat{S}_{(-,n-1)(+,n)}\right)$$

$$+ \sin^2 \varphi_n \cos^2 \varphi_{n-1} \left(2\hat{S}_{(-,n-1)(-,n)} \hat{\rho} \hat{S}^\dagger_{(-,n-1)(-,n)} - \hat{S}^\dagger_{(-,n-1)(-,n)} \hat{S}_{(-,n-1)(-,n)} \hat{\rho} - \hat{\rho} \hat{S}^\dagger_{(-,n-1)(-,n)} \hat{S}_{(-,n-1)(-,n)}\right)\Big).$$

(10)

Here $\gamma_0 = 4\omega_0^3 |\mathbf{d}_{eg}|^2 / 3\hbar c^3$ is the characteristic rate of the transition between eigenstates obtained by averaging over the free space modes (for a detailed derivation, see [13, 19-21]). If we were to consider one atom in free space, then this quantity would be the rate of spontaneous emission.

Equation (10) is obtained after averaging the Hamiltonian of the interaction of the atom with modes of free space over degrees of freedom of modes of free-space. This is rather cumbersome but has a clear physical interpretation. Specifically, each term in Eq. (10) having $2\hat{S}\hat{\rho}\hat{S}^\dagger - \hat{S}^\dagger \hat{S}\hat{\rho} - \hat{\rho}\hat{S}^\dagger \hat{S}$ as a factor describes the transitions of the system between the eigenstates $\{|g,0\rangle, |+,n\rangle, |-,n\rangle\}, n = 1, 2, ...$. The explicit form of the $\hat{S}$-operators is given by Eqs. (9). We emphasize again that it is the interaction of the "atom + selected mode" system with the reservoir of free space modes that leads to a transition between the eigenstates of this subsystem.

Further, we expand the density matrix over eigenstates (2)-(3):

$$\hat{\rho} = \rho_{(g,0)(g,0)} |g,0\rangle\langle g,0| +$$

$$+ \sum_{n=1}^{\infty} \Big(\rho_{(g,0)(+,n)} |g,0\rangle\langle +,n| + \rho_{(g,0)(-,n)} |g,0\rangle\langle -,n| +$$

$$+ \rho_{(+,n)(g,0)} |+,n\rangle\langle g,0| + \rho_{(-,n)(g,0)} |-,n\rangle\langle g,0|\Big)$$

$$+ \sum_{n_1,n_2=1}^{\infty} \Big(\rho_{(+,n_1)(+,n_2)} |+,n_1\rangle\langle +,n_2| + \rho_{(+,n_1)(-,n_2)} |+,n_1\rangle\langle -,n_2|$$

$$+ \rho_{(-,n_1)(+,n_2)} |-,n_1\rangle\langle +,n_2| + \rho_{(-,n_1)(-,n_2)} |-,n_1\rangle\langle -,n_2|\Big).$$

(11)

One of Eq. (10) features is that it defines the independent dynamics of the diagonal and off-diagonal terms of the density matrix [22]. Specifically, from Eq. (10) for diagonal elements, we obtain the following equations

$$\dot{\rho}_{(+,n)(+,n)} = -\gamma_0 \cos^2 \varphi_n \rho_{(+,n)(+,n)} + \gamma_0 \cos^2 \varphi_{n+1} \sin^2 \varphi_n \rho_{(+,n+1)(+,n+1)}$$
$$+\gamma_0 \sin^2 \varphi_{n+1} \sin^2 \varphi_n \rho_{(-,n+1)(-,n+1)}, \tag{12}$$

$$\dot{\rho}_{(-,n)(-,n)} = -\gamma_0 \sin^2 \varphi_n \rho_{(-,n)(-,n)} + \gamma_0 \cos^2 \varphi_{n+1} \cos^2 \varphi_n \rho_{(+,n+1)(+,n+1)}$$
$$+\gamma_0 \sin^2 \varphi_{n+1} \cos^2 \varphi_n \rho_{(-,n+1)(-,n+1)}, \tag{13}$$

$$\dot{\rho}_{(g,0)(g,0)} = \gamma_0 \cos^2 \varphi_1 \rho_{(+,1)(+,1)} + \gamma_0 \sin^2 \varphi_1 \rho_{(-,1)(-,1)}. \tag{14}$$

One can see that for the dynamics of the diagonal elements, only "down" transitions with a decrease in energy are possible. This is a consequence of our assumption that the free space mode reservoir is at zero temperature.

For the off-diagonal elements, we obtain [22]:

$$\dot{\rho}_{(+,n_1)(+,n_2)} = \left(-i\left(\omega_{+,n_2} - \omega_{+,n_1}\right) - \gamma_0 \left(\cos^2 \varphi_{n_2} + \cos^2 \varphi_{n_1}\right)/2\right) \rho_{(+,n_1)(+,n_2)}, \; n_1 \neq n_2. \tag{15}$$

$$\dot{\rho}_{(-,n_1)(+,n_2)} = \left(-i\left(\omega_{+,n_2} - \omega_{-,n_1}\right) - \gamma_0 \left(\cos^2 \varphi_{n_2} + \sin^2 \varphi_{n_1}\right)/2\right) \rho_{(-,n_1)(+,n_2)}, \; n_1 \neq n_2. \tag{16}$$

$$\dot{\rho}_{(+,n_1)(-,n_2)} = \left(-i\left(\omega_{-,n_2} - \omega_{+,n_1}\right) - \gamma_0 \left(\sin^2 \varphi_{n_2} + \cos^2 \varphi_{n_1}\right)/2\right) \rho_{(+,n_1)(-,n_2)}, \; n_1 \neq n_2. \tag{17}$$

$$\dot{\rho}_{(-,n_1)(-,n_2)} = \left(-i\left(\omega_{-,n_2} - \omega_{-,n_1}\right) - \gamma_0 \left(\sin^2 \varphi_{n_2} + \sin^2 \varphi_{n_1}\right)/2\right) \rho_{(-,n_1)(-,n_2)}, \; n_1 \neq n_2. \tag{18}$$

$$\dot{\rho}_{(0,g)(+,n)} = \left(-i\omega_{+,n} - \gamma_0 \cos^2 \varphi_n / 2\right) \rho_{(0,g)(+,n)}. \tag{19}$$

$$\dot{\rho}_{(0,g)(-,n)} = \left(-i\omega_{-,n} - \gamma_0 \sin^2 \varphi_n / 2\right) \rho_{(0,g)(-,n)}. \tag{20}$$

Each off-diagonal element of the density matrix oscillates and simultaneously decays exponentially and independently of other elements.

Suppose that at the initial moment, the whole system is in the state $|g,0\rangle$ with the density matrix $\hat{\rho}(0) = |g,n_0\rangle\langle g,n_0|$. This state is not an eigenstate, and the expansion over eigenstates (2)-(3) gives

$$\hat{\rho}(0) = |g,n_0\rangle\langle g,n_0| = \sin^2 \varphi_{n_0} |n,+\rangle\langle n,+| + \cos^2 \varphi_{n_0} |n,-\rangle\langle n,-|$$
$$+ \sin \varphi_{n_0} \cos \varphi_{n_0} \left(|n_0,+\rangle\langle n_0,-| + |n_0,-\rangle\langle n_0,+|\right) \tag{21}$$

Thus, the initial conditions for Eqs. (12) – (20) have the form:

$$\rho_{(n_0,+)(n_0,+)}(0) = \sin^2 \varphi_{n_0},$$
$$\rho_{(n_0,-)(n_0,-)}(0) = \cos^2 \varphi_{n_0}, \tag{22}$$
$$\rho_{(n_0,-)(n_0,+)} = \rho_{(n_0,+)(n_0,-)}(0) = \sin \varphi_{n_0} \cos \varphi_{n_0},$$

with all other elements of the density matrix equal to zero.

Rayleigh scattering is non-resonant; therefore, for simplicity, we consider the limiting case of a large detuning, $|\Delta| \gg \Omega_R \sqrt{n}$, and a large volume of the resonator, $V \to \infty$. For definiteness, we assume $\omega_{SM} < \omega_{TLS}$. Then, using $\varphi_n = tan^{-1}\left(2\Omega_R \sqrt{n}/|\Delta|\right)/2 \simeq \Omega_R \sqrt{n}/|\Delta|$,

$sin\,\varphi_n \simeq \varphi_n \simeq \Omega_R\sqrt{n}/|\Delta|$, and $cos\,\varphi_n \simeq 1$, the expression for eigenfrequencies (4) and states (3) can be expanded in a small parameter $\Omega_R\sqrt{n}/\Delta$:

$$\omega_{n,-} \approx n\omega_{SM} - \Omega_R^2 n/|\Delta|, \qquad |n,-\rangle \approx -\frac{\Omega_R n}{|\Delta|}|e,n-1\rangle + |g,n\rangle, \qquad (23)$$

$$\omega_{n,+} \approx (n-1)\omega_{SM} + \omega_{TLS} + \Omega_R^2 n/|\Delta|, \qquad |n,+\rangle \approx |e,n-1\rangle + \frac{\Omega_R\sqrt{n}}{|\Delta|}|g,n\rangle. \qquad (24)$$

One can see that in the limit $\Delta \gg \Omega_R\sqrt{n}$, the eigenfrequencies $\omega_{n,-}$, Eq. (23), for the eigenstates $|n,-\rangle$ coincide with the eigenfrequencies $n\omega_{SM}$ of the selected mode with the accuracy $\Omega_R^2 n/|\Delta| \ll 1$, and the difference between the frequencies of the levels $|n,-\rangle$ with different $n$ is equal to $\omega_{CM} - \Omega_R^2/|\Delta|$, i.e., with the accuracy $\Omega_R^2/|\Delta|$ it coincides with the frequency of the selected mode $\omega_{SM}$. Thus, we can expect that a photon with a frequency $\omega_{SM}$ is emitted during the transition between eigenstates $|n,-\rangle$. Note that in this consideration, states with the transition frequency $\omega_{SM}$ arise naturally, and it is not necessary to introduce an additional virtual level artificially.

Let us consider how the initial condition $\hat{\rho}(0) = |g,n_0\rangle\langle g,n_0|$ is expanded over the basis of eigenstates in the limit $\Delta \gg \Omega_R\sqrt{n}$. Using Eqs. (22) and approximate equalities $sin\,\varphi_n \simeq \varphi_n \simeq \Omega_R\sqrt{n}/|\Delta|$ and $cos\,\varphi_n \simeq 1$, we obtain

$$\begin{aligned}\rho_{(-,n_0)(-,n_0)}(0) \simeq 1, \quad \rho_{(+,n_0)(+,n_0)}(0) = \Omega_R^2 n/|\Delta|^2 \ll 1, \\ \rho_{(-,n_0)(+,n_0)}(0) = \rho_{(+,n_0)(-,n_0)}(0) = \Omega_R\sqrt{n}/|\Delta| \ll 1.\end{aligned} \qquad (25)$$

We, therefore, assume that $\rho_{(n_0,-)(n_0,-)}(0) = 1$ and $\rho_{(+,n_0)(+,n_0)}(0) = \rho_{(-,n_0)(+,n_0)}(0) = \rho_{(+,n_0)(-,n_0)}(0) = 0$. As noted above, with a decrease in energy, only downward transitions are realized. Therefore, due to the initial condition $\rho_{(\pm,n)(\pm,n)}(0) = 0$, $n > n_0$, the matrix element $\rho_{(n_0,+)(n_0,+)}$ is equal to zero all the time.

Now, using the approximations obtained above, we consider the diagonal elements of the density matrix. We begin with $\rho_{(n_0,-)(n_0,-)}(t)$. From Eq. (13), using $sin\,\varphi_n \simeq \varphi_n \simeq \Omega_R\sqrt{n}/|\Delta|$ and $cos\,\varphi_n \simeq 1$, we have

$$\dot{\rho}_{(-,n_0)(-,n_0)}(t) = -\gamma_0 \frac{\Omega_R^2 n}{|\Delta|} \rho_{(-,n_0)(-,n_0)}(t). \qquad (26)$$

Since $\rho_{(+,n_0)(+,n_0)}(0) = 0$, from Eq. (13) we obtain

$$\dot{\rho}_{(-,n_0-1)(-,n_0-1)}(t) = -\gamma_0 \frac{\Omega_R^2 (n-1)}{|\Delta|} \rho_{(-,n_0-1)(-,n_0-1)}(t) + \gamma_0 \frac{\Omega_R^2 n}{|\Delta|} \rho_{(-,n_0)(-,n_0)}(t). \qquad (27)$$

Next, we consider the equation for $\rho_{(+,n_0-1)(+,n_0-1)}(t)$. From Eq. (12) and using $\rho_{(+,n_0)(+,n_0)}(t) = 0$ we obtain

$$\dot{\rho}_{(+,n_0-1)(+,n_0-1)}(t) = -\gamma_0 \rho_{(+,n_0-1)(+,n_0-1)}(t) + \gamma_0 \frac{\Omega_R^4 n(n-1)}{|\Delta|^4} \rho_{(-,n_0)(-,n_0)}(t),$$

$$\rho_{(+,n_0-1)(+,n_0-1)}(0) = 0. \tag{28}$$

The factor $\Omega_R^4 n(n-1)/|\Delta|^4$ is of the second order of smallness with respect to the parameter $\Omega_R^2 n / |\Delta| \ll 1$. Therefore, in Eq. (28), we neglect the term $\left(\gamma_0 \Omega_R^4 n(n-1)/|\Delta|^4\right)\rho_{(-,n_0)(-,n_0)}$. Then, solution (28) is $\rho_{(+,n_0-1)(+,n_0-1)}(t) = 0$.

Repeating the above arguments for the elements $\rho_{(+,n)(+,n)}$, $n < n_0 - 1$, of the density matrix, we obtain that $\rho_{(+,n)(+,n)}(t) = 0$. At the same time, for $\rho_{(-,n)(-,n)}(t)$, $n < n_0 - 1$ we obtain the equation of the same form as Eq. (27). For brevity, let us denote $\rho_{(n,-)(n,-)}(t) \equiv p_n(t)$, $\gamma_0 \Omega_R^2 n / |\Delta| \equiv \gamma_n$, and $\omega_{n,-} = \omega_n$. The quantity $p_n(t)$ has the meaning of the probability of the occupation of the entangled state of the selected mode and the atom $|-, n\rangle = -\sin\varphi_n |e, n-1\rangle + \cos\varphi_n |g, n\rangle$, while $\gamma_n$ has the meaning of the transition rate form the state $|-, n\rangle$ to the state $|-, n-1\rangle$. Equations for $p_n(t)$ take the form:

$$\begin{aligned}\dot{p}_{n_0}(t) &= -\gamma_{n_0} p_{n_0} \\ \dot{p}_n(t) &= -\gamma_n p_n(t) + \gamma_{n+1} p_{n+1}(t), \quad 1 \leq n < n_0 \\ \dot{p}_0(t) &= \gamma_1 p_1(t)\end{aligned} \tag{29}$$

with the initial condition

$$p_{n_0}(0) = 1, \ p_n(0) = 0, \ 0 \leq n < n_0. \tag{30}$$

According to the Wiener-Khinchin theorem [17, 23, 24], the scattering spectrum can be expressed in terms of the Fourier transform of the two-time correlation function $\langle \hat{\sigma}^\dagger(t_1)\hat{\sigma}(t_2)\rangle$, $t_1 > t_2$ [13, 19], where the operator $\hat{\sigma}$ in the basis of eigenstates has the form of Eq. (7). According to the general theory [13, 17, 19], to find $\langle \hat{\sigma}^\dagger(t_1)\hat{\sigma}(t_2)\rangle$, $t_1 > t_2$, one needs to solve master equation (10) two times with two different initial conditions. First, we need to solve Eq. (10) with initial condition (21), and find $\hat{\rho}(t_2)$. Then, it is necessary to solve master equation (10) for $t \leq t_1$ with the initial condition $\hat{\rho}(0) = \hat{\sigma}\hat{\rho}(t_2)$. The obtained value we denote as $\hat{\tilde{\rho}}(t_1 - t_2)$. According to the quantum regression theorem [13, 17, 19], $\langle \hat{\sigma}^\dagger(t_1)\hat{\sigma}(t_2)\rangle$ is equal to $\text{Tr}\left(\hat{\sigma}^\dagger \hat{\tilde{\rho}}(t_1 - t_2)\right)$.

In terms of the solution of system (29) with initial conditions (30), the one-time average of the atom population can be written as

$$\langle \hat{\sigma}^{\dagger}(t)\hat{\sigma}(t) \rangle \simeq \sum_{k=n_0}^{1} \sin^2 \varphi_k p_k(t) \simeq \sum_{k=n_0}^{1} \left( \Omega_R^2 k / |\Delta|^2 \right) p_k(t), \qquad (31)$$

while the two-time averages, with the aid of Eqs. (15)–(20), can be rewritten in the form

$$\langle \hat{\sigma}^{\dagger}(t+\tau)\hat{\sigma}(t) \rangle \simeq \sum_{k=n_0}^{1} \left( \Omega_R^2 k / |\Delta| \right)^2 p_k(t) \exp\left( \left( i\omega_{CM} - (\gamma_k + \gamma_{k-1})/2 \right) \tau \right). \qquad (32)$$

A specific calculation of the radiation depends on the initial value of the number of photons $n_0$ in the selected mode, and how we obtain the limit $V \to \infty$. Different limits give the answers obtained within the approaches of Placzek [8], Berestetskii-Lifshitz-Pitaevskii [6], and the coherent initial state [25] suggested by Glauber. The latter case, in our opinion, corresponds most closely to the formulation of the problem in which the field is considered as classical.

Thus, we develop a theory in which scattering is considered as a relaxation of an interacting atom and a photon from the selected mode to the reservoir of the other free space modes. We obtain the master equation for the system dynamics, Eq. (10), and simplify this equation in the case of large detuning between the frequency of selected mode and atomic transition frequency, Eqs. (29). Note that all the levels in the system considered, Eq. (4), are real, and there is no necessity to introduce a virtual level. Below, we show that the model developed reproduces the elasticity of the Rayleigh scattering, and in the case of a coherent initial state of a photon from the selected mode, it yields the same results as a theory in which an incident photon is described as a classical field.

## 3. Single-photon scattering

Strictly speaking, to solve the problem of scattering of $n_0$ photons, it is necessary to solve system of equations (29), which takes into account the transitions between all eigenstates (23) (it is solved in the next section). However, usually, we only need to consider a transition in which the number of photons in the selected mode changes by one. We do this in this section within the framework of the theory developed above.

*3.1 The Berestetskii-Lifshitz-Pitaevskii approach of a single incident photon*

First, we assume that initially in the selected mode, only one photon is present, i.e., the initial condition is $n_0 = 1$. Such a formulation was considered in Ref. [6].

As mentioned above, without interaction with the reservoir, in the system consisting of the atom and photon, the atom's probability of being excited, $\langle \hat{\sigma}^{\dagger}(t)\hat{\sigma}(t) \rangle \equiv p_e(t)$, and the mean number of photons, $\langle \hat{a}^{\dagger}(t)\hat{a}(t) \rangle \equiv n_{ph}(t)$, should exhibit Rabi oscillations. The interaction of the system with the free space mode reservoir leads to vanishing with the rate $\gamma_1$ from the Eq. (29) of both $p_e(t)$ and $n_{ph}(t)$ against the background of damping of the Rabi oscillations. Indeed, numerical simulation of Eq. (10) demonstrates such behavior (see Fig. 4). From this numerical simulation, we conclude that there are two characteristic times in the system dynamics, $\gamma_0^{-1}$ and $\left(\gamma_0 \Omega_R^2 / |\Delta|^2\right)^{-1}$.

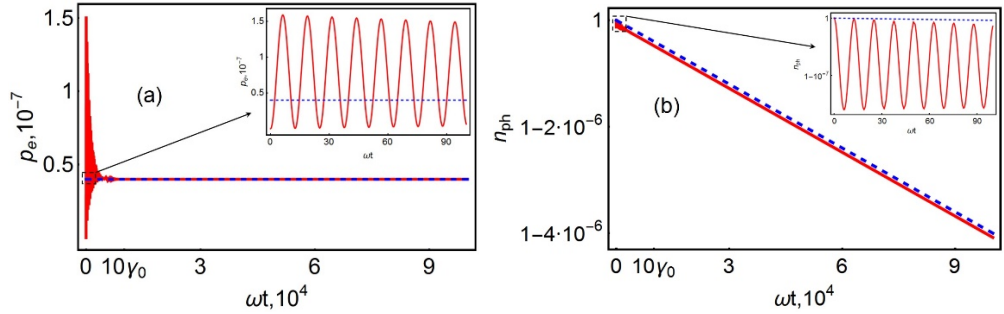

Fig. 1. Time dependence of (a) the probability of the atom to be in the excited state and (b) the number of photons in the selected mode.

We can see that at the initial time, $t \leq \gamma_0^{-1}$, both $p_e(t)$ and $n_{ph}(t)$ oscillate with the Rabi frequency $2\sqrt{\Omega_R^2 + (\Delta/2)^2}$. The Rabi oscillations of both quantities dissipate with the characteristic rate $\sim \gamma_0$. The presence of the Rabi oscillations means that the photon from the selected mode does not absorb.

After the dissipation of Rabi oscillations, the probability of the atom to be in the excited state reaches the value $\left(\Omega_R^2 / |\Delta|^2\right)$. At the time $t \gg \gamma_0^{-1}$, the system dynamics obeys Eq. (29), which solution has the form

$$p_1(t) = exp\left(-\gamma_0 \left(\Omega_R^2 / |\Delta|\right) t\right). \tag{33}$$

The solution for the atom population, Eq. (31), is

$$\langle \hat{\sigma}^\dagger(t)\hat{\sigma}(t)\rangle \equiv p_e(t) \simeq \left(\Omega_R^2 / |\Delta|^2\right) exp\left(-\gamma_0 \left(\Omega_R^2 / |\Delta|^2\right) t\right). \tag{34}$$

Equation (34) shows that the atom radiates with the rate $\gamma_0 \left(\Omega_R^2 / |\Delta|^2\right)$. At the initial time, the excitation energy is $\Omega_R^2 / |\Delta|^2$. Usually, in problems about an atom in free space, the infinite volume limit is considered. In this limit, the coupling constant approaches zero as $\Omega_R \sim 1/\sqrt{V} \to 0$ and the radiation rate is infinitely small. In this formulation, two-time correlation function (32) has the form

$$\langle \hat{\sigma}^\dagger(t+\tau)\hat{\sigma}(t)\rangle \simeq exp(i\omega_{SM}\tau)\Omega_R^2 / |\Delta|^2 exp\left(-\gamma_0 \left(\Omega_R^2 / |\Delta|\right) t\right) exp\left(-\gamma_0 \left(\Omega_R^2 / 2|\Delta|^2\right)\tau\right). \tag{35}$$

Note that if we fix the time and find the spectrum as the Fourier transform of the two-time correlation function with respect to $\tau$, then we obtain a Lorentz line centered at the frequency $\omega_{SM}$ of the selected mode with the width $\sim \gamma_0 \left( \Omega_R^2 / 2|\Delta|^2 \right) \ll \gamma_0$. Thus, in this formulation, a narrow emission line is reproduced. However, this narrow line corresponds to extremely slow radiation. Further, the total detectable energy emitted by the atom is $\sim \Omega_R^2 / |\Delta|^2$, i.e., in the limit considered, it is also extremely small.

One can calculate the scattering cross section by dividing the total radiated energy by the energy flux that the selected mode creates at the atom location. Since the energy flux $\sim \Omega_R^2$ [17, 26], small quantities are canceled out, and the cross section becomes finite (it is given in Refs. [8] and [6]). However, the finite value of the scattering cross section is obtained due to the division of two infinitely small quantities: incident and radiated energies. Such problem formulation does not correspond to an experiment in which the finite field energy flows over a finite time, and the final radiation energy is detected.

The smallness of the radiation energy can be avoided if we consider the limit at which the average field value at the atom location is finite. For this, we put $\Omega_R \sim 1/\sqrt{V} \to 0$ and $n_0 \sim V \to \infty$, then $\Omega_R \sqrt{n_0} \to const$, i.e., we consider a large initial value of the number of photons in the selected mode so that for any volume, the electric field of the selected mode is finite. We continue to assume that the detuning is so large that $\Omega_R \sqrt{n_0} / \Delta \ll 1$.

*3.2. Placzek's approach. The scattering of one photon from ensemble of photons*

In the approach developed by Placzek [8], one considers a single transition process, in which the number of photons is changed by unity. To reproduce the result of this approach, it is necessary, in system (29), to retain only one equation that describes the dynamics of the initial state $|n_0, -\rangle$. We assume that the initial state of the system is $p_{n_0}(0) = 1$. Other states are supposed to be empty all the time, i.e. $p_n(0) = p_n(t) = 0, 1 \le n < n_0$. After such simplification, the system of Eqs. (29) reduces to

$$\dot{p}_{n_0}(t) = -\gamma_{n_0} p_{n_0} \qquad (36)$$

Solving this equation for the population of the excited state of the atom, we obtain

$$\langle \hat{\sigma}^\dagger(t)\hat{\sigma}(t) \rangle \simeq \left( \Omega_R^2 n / |\Delta|^2 \right) exp\left( -\gamma_0 \left( \Omega_R^2 n / |\Delta|^2 \right) t \right). \qquad (37)$$

For the two-time correlation function, we have

$$\langle \hat{\sigma}^\dagger(t+\tau)\hat{\sigma}(t) \rangle \simeq exp(i\omega_{SM}\tau) \Omega_R^2 n / |\Delta|^2 \, exp\left( -\gamma_0 \left( \Omega_R^2 n / |\Delta| \right) t \right) exp\left( -\gamma_0 \left( \Omega_R^2 n / 2|\Delta|^2 \right) \tau \right). \qquad (38)$$

If we fix time and make the Fourier transform of Eq. (38), we obtain the Lorentzian line with the width $\gamma_0 \left( \Omega_R^2 n / 2|\Delta|^2 \right) \ll \gamma_0$. In this case, the total radiated energy is finite and equal $\Omega_R^2 n / |\Delta|^2$.

Thus, in the limit $\Omega_R \sim 1/\sqrt{V} \to 0$, $n_0 \sim V \to \infty$, and $\Omega_R \sqrt{n_0} \to const$, the result is close to experiment: the spectrum is centered on the frequency of the selected mode $\omega_{SM}$ and has a narrow line $\gamma_0 \left( \Omega_R^2 n / 2|\Delta|^2 \right) \ll \gamma_0$.

However, the important difference from experiment and from the formulation of the problem with the classical field remains. This is the choice of the time $t$, after which one starts measuring the spectrum. In experiment and in the formulation of the problem with a classical field, the population of the excited state of an atom reaches a stationary state, and the time $t$ can formally be set to infinity. The system reaches dynamic equilibrium: at each moment of time, some energy is radiated and some enters from the external field. However, in the above solution, if $t$ tends to infinity, the factor $exp\left(-\gamma_0 \left( \Omega_R^2 n / |\Delta| \right) t \right)$ results in zero probability of an atom to be in the excited state, Eq. (38). This corresponds to the fact that the system from the state with $n_0$ photons passes to a state with $n_0 - 1$ photons - one photon is emitted then the process stops. However, as indicated above, it is necessary to consider subsequent processes of transition between eigenstates and photon radiation. This is done in the next section.

## 4. Many photons in the mode, multiple transitions between the eigenstates of the "selected mode + atom" system. Achieving dynamic equilibrium

Consider the same limit as in the previous case, $\Omega_R \sim 1/\sqrt{V} \to 0$, $n_0 \sim V \to \infty$, and $\Omega_R \sqrt{n_0} \to const$. In the previous section, we take into account only the transition between the states $|n_0, -\rangle$ and $|n_0 - 1, -\rangle$, and only solve equations for $p_{n_0}$ and $p_{n_0-1}$ from system (29). Now we solve full system of equations (29). In the case $n_0 \gg 1$, the rate of transitions between states for which $|n - n_0| \ll n_0$ are approximately the same: $\gamma_n = \gamma_0 \left( \Omega_R^2 n / |\Delta|^2 \right) \approx \gamma_0 \left( \Omega_R^2 n_0 / |\Delta|^2 \right) \equiv \gamma_{eff}$.
System (29), in which all transition rates are the same, corresponds to the Markov-Poisson process, which solution is well known [27]:

$$p_{n_0}(t) = exp(-\gamma_{eff} t),$$
$$p_{n_0}(t) = \left( \gamma_{eff} t / 1! \right)^1 exp(-\gamma_{eff} t), \qquad (39)$$
$$p_{n_0-k}(t) = \left( \gamma_{eff} t / k! \right)^k exp(-\gamma_{eff} t).$$

Using solution (39) we find the average population of the excited state of the atom

$$\left\langle \hat{\sigma}^\dagger(t) \hat{\sigma}(t) \right\rangle = \left( \Omega_R^2 n_0 / |\Delta|^2 \right) exp(-\gamma_{eff} t) \sum_{k=n_0}^{1} \frac{\left( \gamma_{eff} t \right)^{n_0 - k}}{k!} \xrightarrow[n_0 \to \infty]{} \left( \Omega_R^2 n_0 / |\Delta|^2 \right). \qquad (40)$$

One can see that in the limit of a large initial number of photons in the selected mode, the population of the upper state of the atom reaches the constant value of $\left( \Omega_R^2 n_0 / |\Delta|^2 \right)$. Note that this value is proportional to the square of the dipole moment, $\Omega_R^2 \sim \omega_{CM} d^2 / \hbar V$ which is obtained in the problem with the classical field [19].

For a two-time correlation function, we obtain

$$\langle \hat{\sigma}^\dagger(t+\tau)\hat{\sigma}(t) \rangle \simeq$$
$$\simeq exp(i\omega_{SM}\tau) exp\left(-\gamma_0\left(\Omega_R^2 n / |\Delta|^2\right)\tau\right) \Omega_R^2 n_0 / |\Delta|^2 \, exp\left(-\gamma_0\left(\Omega_R^2 n / |\Delta|\right)t\right) \sum_{k=n_0}^{1} \frac{(\gamma_{eff} t)^{n_0 - k}}{k!}. \qquad (41)$$

In the limit $n_0 \sim V \to \infty$, $t \to \infty$, and $\gamma_{eff} t / n_0 \to 0$, we have

$$\langle \hat{\sigma}^\dagger(t_{st}+\tau)\hat{\sigma}(t_{st}) \rangle \simeq \frac{\Omega_R^2 n_0}{\Delta^2} exp\left(\left(i\omega_{CM} - \gamma_0 \frac{\Omega_R^2 n_0}{\Delta^2}\right)\tau\right). \qquad (42)$$

The limit $\gamma_{eff} t / n_0 \to 0$ reflects that during the system evolution, the time is such that the initial number of photons in the selected mode does not deviate significantly from $n_0$.

The two-time correlation function corresponds to the spectrum that is centered on the frequency of the selected mode and has the width $\gamma_0(\Omega_R^2 n_0 / \Delta^2) \ll \gamma_0$. We also note that the two-time correlation function is calculated after the time $t_{st}$ for which the population of the excited state reaches the stationary value $(\Omega_R^2 n_0 / \Delta^2)$. This corresponds to the spectrum that is measured in a system that has reached dynamic equilibrium.

For the radiation spectrum, we obtain (see Ref. [19])

$$\langle \hat{a}_{\mathbf{k},\lambda}^\dagger(t)\hat{a}_{\mathbf{k},\lambda}(t) \rangle \equiv n_{\mathbf{k},\lambda}(t) =$$
$$= \left|\sqrt{\frac{2\pi\omega_\mathbf{k}}{V\hbar}} \mathbf{e}_{\mathbf{k},\lambda} \cdot \mathbf{d}_{eg}\right|^2 \int_0^t d\tau_2 \int_0^t d\tau_1 \langle \hat{\sigma}^\dagger(\tau_2)\hat{\sigma}(\tau_1) \rangle exp(-i\omega_{\mathbf{k},\lambda}(\tau_2 - \tau_1))$$
$$\simeq \left|\sqrt{\frac{2\pi\omega_\mathbf{k}}{V\hbar}} \mathbf{e}_{\mathbf{k},\lambda} \cdot \mathbf{d}_{eg}\right|^2 t_{st} \pi \int_0^\infty d\tau \langle \hat{\sigma}^\dagger(t_{st}+\tau)\hat{\sigma}(t_{st}) \rangle exp(-i\omega_{\mathbf{k},\lambda}\tau) \qquad (43)$$
$$\simeq \frac{\left|\sqrt{\frac{2\pi\omega_\mathbf{k}}{V\hbar}} \mathbf{e}_{\alpha,\mathbf{k}} \cdot \mathbf{d}_{eg}\right|^2 t_{st} \left(\gamma_0 \Omega_R^2 n_0 / \Delta^2\right)^2}{(\omega_{\mathbf{k},\lambda} - \omega_{SM})^2 + (\gamma_0 \Omega_R^2 n_0 / \Delta^2)^2}$$

Thus, we obtain the Lorentzian distribution of photons in frequencies, centered on the frequency of the selected mode and having the width $\gamma_0 \Omega_R^2 n_0 / \Delta^2 \ll \gamma_0$. The number of photons increases with time. This reflects that the system has reached dynamic equilibrium and continuously emits photons.

Let us compare the dynamics of the process considered here with the dynamics in Placzek's theory. In the latter, it was assumed the system transitions from the initial state to a given final state. Subsequent transitions were not considered. In the approach developed here, the system make a cascade of transitions between the entangled states $|-,n\rangle$, until the system reaches its ground state $|g,0\rangle$.

## 5. Coherent state of the mode. Formation of the atomic dipole moment. Coherent Rayleigh scattering

In the previous sections, we show that when all possible transitions between the eigenstates are taken into account, the population inversion of the atom reaches a quasistationary value that does not change until a large number of photons remains in the selected mode. However, there is still a discrepancy with the classical formulation of the problem. Since the off-diagonal elements of the density matrix are equal to zero [see Eq. (22)], the average value of the atomic dipole moment is also zero,

$$\langle \hat{\sigma}(t) \rangle = \cos\varphi_1 \rho_{(g,0)(+,1)} - \sin\varphi_1 \rho_{(g,0)(-,1)} + \sum_{n=2}^{\infty}(\cos\varphi_n \sin\varphi_{n-1}\rho_{(+,n-1)(+,n)}$$
$$- \sin\varphi_n \sin\varphi_{n-1}\rho_{(+,n-1)(-,n)} + \cos\varphi_n \cos\varphi_{n-1}\rho_{(-,n-1)(+,n)} - \sin\varphi_n \cos\varphi_{n-1}\rho_{(-,n-1)(-,n)}) = 0.$$
(44)

Note that this equation differs significantly from the classical case. When the field is described classically, the dipole moment is nonzero, and it oscillates with the frequency of the external field. Thus, Eq. (44) corresponds to incoherent light. An experimentally realized situation can be described adequately by a theory in which light is described classically. Therefore, for the field, one has to choose the initial condition in which a nonzero dipole moment is excited.

For this, it is necessary to consider the coherent state of the field as the initial condition, i.e., instead of $\hat{\rho}(0) = |g,n_0\rangle\langle g,n_0|$, take $\hat{\rho}(0) = |g,\alpha\rangle\langle g,\alpha|$, where $|\alpha\rangle = \exp\left(-|\alpha|^2/2\right)\sum_{n=0}^{\infty}\frac{\alpha^n}{\sqrt{n!}}|n\rangle$ is the coherent state of the field. This state has a nonzero average value of electric and magnetic fields: $\langle \alpha|\hat{E}(\mathbf{r})|\alpha\rangle = \sqrt{2\pi\hbar\omega_{SM}/V}\, Re\,\alpha\,\exp(i\mathbf{k}\mathbf{r})$ [13]. The expansion of such an initial state over eigenstates (3) has the form

$$\hat{\rho}(0) = \exp\left(-|\alpha|^2\right)\Big[|g,0\rangle\langle g,0| +$$
$$+ \sum_{n_1=1}^{\infty}\left(\sin\varphi_{n_1}\frac{\alpha^{n_1}}{\sqrt{n_1!}}|+,n_1\rangle\langle g,0| + \cos\varphi_{n_1}\frac{\alpha^{n_1}}{\sqrt{n_1!}}|-,n_1\rangle\langle g,0|\right) +$$
$$+ \sum_{n_2=1}^{\infty}\left(\sin\varphi_{n_2}\frac{\alpha^{*n_2}}{\sqrt{n_2!}}|g,0\rangle\langle +,n_2| + \cos\varphi_{n_2}\frac{\alpha^{*n_2}}{\sqrt{n_2!}}|g,0\rangle\langle -,n_2|\right) +$$
(45)
$$+ \sum_{n_1,n_2=1}^{\infty}\left(\sin\varphi_{n_1}\sin\varphi_{n_2}\frac{\alpha^{n_1}}{\sqrt{n_1!}}\frac{\alpha^{*n_2}}{\sqrt{n_2!}}|+,n_1\rangle\langle +,n_2| + \sin\varphi_{n_1}\cos\varphi_{n_2}\frac{\alpha^{n_1}}{\sqrt{n_1!}}\frac{\alpha^{*n_2}}{\sqrt{n_2!}}|+,n_1\rangle\langle -,n_2| +$$
$$+ \cos\varphi_{n_1}\sin\varphi_{n_2}\frac{\alpha^{n_1}}{\sqrt{n_1!}}\frac{\alpha^{*n_2}}{\sqrt{n_2!}}|-,n_1\rangle\langle +,n_2| + \cos\varphi_{n_1}\cos\varphi_{n_2}\frac{\alpha^{n_1}}{\sqrt{n_1!}}\frac{\alpha^{*n_2}}{\sqrt{n_2!}}|-,n_1\rangle\langle -,n_2|\right)\Big].$$

In expansion (45) of the density matrix over eigenstates, in contrast to expansion (21), there are all kinds of off-diagonal elements. Equation (44) shows that to calculate the average dipole moment, we need to know the dynamics of not all off-diagonal matrix elements, but only $\rho_{(g,0)(+,1)}$, $\rho_{(+,n-1)(+,n)}$, $\rho_{(-,n-1)(+,n)}$, $\rho_{(-,n-1)(-,n)}$. According to Eqs. (15)–(20), these matrix elements obey the equations

$$\dot{\rho}_{(+,n-1)(+,n)} = \left(-i(\omega_{+,n} - \omega_{+,n-1}) - \frac{\gamma_0}{2}(\cos^2\varphi_n + \cos^2\varphi_{n-1})\right)\rho_{(+,n-1)(+,n)}, \tag{46}$$

$$\dot{\rho}_{(-,n-1)(+,n)} = \left(-i(\omega_{+,n} - \omega_{-,n-1}) - \frac{\gamma_0}{2}(\cos^2\varphi_n + \sin^2\varphi_{n-1})\right)\rho_{(-,n-1)(+,n)}, \tag{47}$$

$$\dot{\rho}_{(+,n-1)(-,n)} = \left(-i(\omega_{-,n} - \omega_{+,n-1}) - \frac{\gamma_0}{2}(\sin^2\varphi_n + \cos^2\varphi_{n-1})\right)\rho_{(+,n-1)(-,n)}, \tag{48}$$

$$\dot{\rho}_{(-,n-1)(-,n)} = \left(-i(\omega_{-,n} - \omega_{-,n-1}) - \frac{\gamma_0}{2}(\sin^2\varphi_n + \sin^2\varphi_{n-1})\right)\rho_{(-,n-1)(-,n)}, \tag{49}$$

$$\dot{\rho}_{(0,g)(+,1)} = \left(-i\omega_{+,1} - \gamma_0 \cos^2\varphi_1/2\right)\rho_{(0,g)(+,1)}, \tag{50}$$

$$\dot{\rho}_{(0,g)(-,1)} = \left(-i\omega_{-,1} - \gamma_0 \sin^2\varphi_1/2\right)\rho_{(0,g)(-,1)}, \tag{51}$$

and in accordance with Eq. (45), the initial conditions of these matrix elements have the form

$$\rho_{(+,n-1)(+,n)}(0) = \exp(-|\alpha|^2)\sin\varphi_n \sin\varphi_{n-1}\frac{\alpha^{n-1}}{\sqrt{(n-1)!}}\frac{\alpha^{*n}}{\sqrt{n!}}, \tag{52}$$

$$\rho_{(-,n-1)(+,n)}(0) = \exp(-|\alpha|^2)\cos\varphi_n \sin\varphi_{n-1}\frac{\alpha^{n-1}}{\sqrt{(n-1)!}}\frac{\alpha^{*n}}{\sqrt{n!}}, \tag{53}$$

$$\rho_{(+,n-1)(-,n)}(0) = \exp(-|\alpha|^2)\sin\varphi_n \cos\varphi_{n-1}\frac{\alpha^{n-1}}{\sqrt{(n-1)!}}\frac{\alpha^{*n}}{\sqrt{n!}}, \tag{54}$$

$$\rho_{(-,n-1)(-,n)}(0) = \exp(-|\alpha|^2)\cos\varphi_n \cos\varphi_{n-1}\frac{\alpha^{n-1}}{\sqrt{(n-1)!}}\frac{\alpha^{*n}}{\sqrt{n!}}, \tag{55}$$

$$\rho_{(0,g)(+,1)}(0) = \exp(-|\alpha|^2)\sin\varphi_1\frac{\alpha^*}{\sqrt{1!}}, \tag{56}$$

$$\rho_{(0,g)(-,1)}(0) = \exp(-|\alpha|^2)\cos\varphi_1\frac{\alpha^*}{\sqrt{1!}}. \tag{57}$$

Solutions (46)–(51) with initial conditions (52)–(57) are

$$\rho_{(+,n-1)(+,n)}(t) = e^{-|\alpha|^2}\sin\varphi_n \sin\varphi_{n-1}\frac{\alpha^{n-1}}{\sqrt{(n-1)!}}\frac{\alpha^{*n}}{\sqrt{n!}}e^{\left(-i(\omega_{+,n}-\omega_{+,n-1})-\frac{\gamma_0}{2}(\cos^2\varphi_n+\cos^2\varphi_{n-1})\right)t}, \tag{58}$$

$$\rho_{(-,n-1)(+,n)}(t) = e^{-|\alpha|^2}\cos\varphi_n \sin\varphi_{n-1}\frac{\alpha^{n-1}}{\sqrt{(n-1)!}}\frac{\alpha^{*n}}{\sqrt{n!}}e^{\left(-i(\omega_{+,n}-\omega_{-,n-1})-\frac{\gamma_0}{2}(\cos^2\varphi_n+\sin^2\varphi_{n-1})\right)t}, \tag{59}$$

$$\rho_{(+,n-1)(-,n)}(t) = e^{-|\alpha|^2} \sin\varphi_n \cos\varphi_{n-1} \frac{\alpha^{n-1}}{\sqrt{(n-1)!}} \frac{\alpha^{*n}}{\sqrt{n!}} e^{\left(-i(\omega_{-,n}-\omega_{+,n-1}) - \frac{\gamma_0}{2}\left(\sin^2\varphi_n + \cos^2\varphi_{n-1}\right)\right)t}, \tag{60}$$

$$\rho_{(-,n-1)(-,n)}(t) = e^{-|\alpha|^2} \cos\varphi_n \cos\varphi_{n-1} \frac{\alpha^{n-1}}{\sqrt{(n-1)!}} \frac{\alpha^{*n}}{\sqrt{n!}} e^{\left(-i(\omega_{-,n}-\omega_{-,n-1}) - \frac{\gamma_0}{2}\left(\sin^2\varphi_n + \sin^2\varphi_{n-1}\right)\right)t}, \tag{61}$$

$$\rho_{(0,g)(+,1)}(t) = e^{-|\alpha|^2} \sin\varphi_1 \frac{\alpha^*}{\sqrt{1!}} e^{\left(-i\omega_{+,1} - \gamma_0 \cos^2\varphi_1/2\right)t}, \tag{62}$$

$$\rho_{(0,g)(-,1)}(t) = e^{-|\alpha|^2} \cos\varphi_1 \frac{\alpha^*}{\sqrt{1!}} e^{\left(-i\omega_{-,1} - \gamma_0 \sin^2\varphi_1/2\right)t}. \tag{63}$$

Substituting Eqs. (58)–(63) into the expression for the dipole moment, Eq. (44), we obtain

$$\langle \hat{\sigma}(t) \rangle = e^{-|\alpha|^2} \left( \cos\varphi_1 \sin\varphi_1 \frac{\alpha^*}{\sqrt{1!}} e^{\left(-i\omega_{+,1} - \gamma_0 \cos^2\varphi_1/2\right)t} - \sin\varphi_1 \cos\varphi_1 \frac{\alpha^*}{\sqrt{1!}} e^{\left(-i\omega_{-,1} - \gamma_0 \sin^2\varphi_1/2\right)t} \right.$$

$$+ \sum_{n=2}^{\infty} \frac{\alpha^{n-1}}{\sqrt{(n-1)!}} \frac{\alpha^{*n}}{\sqrt{n!}} \left( \cos\varphi_n \sin^2\varphi_{n-1} \sin\varphi_n e^{\left(-i(\omega_{+,n}-\omega_{+,n-1}) - \frac{\gamma_0}{2}(\cos^2\varphi_n + \cos^2\varphi_{n-1})\right)t} \right.$$

$$- \sin^2\varphi_n \sin\varphi_{n-1} \cos\varphi_{n-1} e^{\left(-i(\omega_{-,n}-\omega_{+,n-1}) - \frac{\gamma_0}{2}(\sin^2\varphi_n + \cos^2\varphi_{n-1})\right)t} \tag{64}$$

$$+ \cos^2\varphi_n \cos\varphi_{n-1} \sin\varphi_{n-1} e^{\left(-i(\omega_{+,n}-\omega_{-,n-1}) - \frac{\gamma_0}{2}(\cos^2\varphi_n + \sin^2\varphi_{n-1})\right)t}$$

$$\left. \left. - \sin\varphi_n \cos^2\varphi_{n-1} \cos\varphi_n e^{\left(-i(\omega_{-,n}-\omega_{-,n-1}) - \frac{\gamma_0}{2}(\sin^2\varphi_n + \sin^2\varphi_{n-1})\right)t} \right) \right).$$

In the limit of a large detuning, $\Omega_R^2 |\alpha|^2 / \Delta^2 \ll 1$, the quantities $\Omega_R^2 |\alpha|^2 / \Delta^2$ and $\sin\varphi_n \simeq \Omega_R \sqrt{n} / |\Delta|$ are small parameters, while $\cos\varphi_n \simeq 1$. Therefore, the main contribution to sum (64) is made by the first, second, fifth, and sixth terms, which are proportional to the first power of the small parameter. Further, the decay rate of the first and fifth terms are $\gamma_0 \cos^2\varphi_1 / 2 \simeq \gamma_0 / 2$ and $\gamma_0 \left(\cos^2\varphi_n + \sin^2\varphi_{n-1}\right)/2 \simeq \gamma_0 / 2$, respectively. Thus, on the time-scale $t \gg \gamma_0^{-1}$, these terms are equal to zero. In turn, the decay rate of the second and six terms are $\gamma_0 \sin^2\varphi_1 / 2 \simeq \gamma_0 \Omega_R^2 n / 2|\Delta|^2 \ll \gamma_0$ and $\gamma_0 \left(\sin^2\varphi_n + \sin^2\varphi_{n-1}\right)/2 \simeq \gamma_0 \Omega_R^2 n / |\Delta|^2 \ll \gamma_0$, respectively. Thus, we conclude that on the time-scale $t \gg \gamma_0^{-1}$, the second and sixth terms give the main contribution to Eq. (64):

$$\langle \hat{\sigma}(t) \rangle = -\frac{e^{-i\omega_{SM} t} \Omega_R}{|\Delta|} e^{-|\alpha|^2} \left( \frac{\alpha^*}{\sqrt{1!}} e^{-\left(\gamma_0 \Omega_R^2 n/2|\Delta|^2\right)t} + \sum_{n=2}^{\infty} \sqrt{n} \frac{\alpha^{n-1}}{\sqrt{(n-1)!}} \frac{\alpha^{*n}}{\sqrt{n!}} e^{-\left(\gamma_0 \Omega_R^2 n/|\Delta|^2\right)t} \right), \tag{65}$$

where expression (23) is used for eigenfrequencies.

Now, assuming that the condition for the non-resonant excitation $\Omega_R^2 |\alpha|^2 / \Delta^2 \ll 1$ remains true, we consider the limit of a large amplitude of the exciting field, $|\alpha|^2 \gg 1$. Since in this case, the main contribution to sum (65) is made by the terms for which $n \simeq |\alpha|^2$, we can calculate this sum approximately. Assuming that inside the sum, the value $\sqrt{n}$ is slowly changing and replacing it with the value $\sqrt{n} \simeq |\alpha|$, we obtain:

$$\langle \hat{\sigma}(t) \rangle \simeq -\frac{e^{-i\omega_{SM}t}\Omega_R |\alpha|}{|\Delta|} e^{-|\alpha|^2} \sum_{n=0}^{\infty} \frac{|\alpha|^{2n}}{n!} e^{-(\gamma_0 \Omega_R^2 n/|\Delta|^2)t} = \frac{e^{-i\omega_{SM}t}\Omega_R |\alpha|}{|\Delta|} exp\left(|\alpha|^2 \left(e^{-(\gamma_0 \Omega_R^2/|\Delta|^2)t} - 1\right)\right). \tag{66}$$

Next, we consider times at which the field amplitude in the selected mode remains almost unchanged, i.e., times $t \ll \left(\gamma_0 \Omega_R^2 / |\Delta|^2\right)^{-1}$. Then, expression (66) is simplified:

$$\langle \hat{\sigma}(t) \rangle \simeq \frac{e^{-i\omega_{SM}t}\Omega_R |\alpha|}{|\Delta|} exp\left(-\frac{|\alpha|^2 \gamma_0 \Omega_R^2}{|\Delta|^2} t\right). \tag{67}$$

Equation (67) describes quasistationary oscillations of a dipole with the frequency of the excited mode. The spectrum of the signal described by Eq. (67) represents the Lorentz line centered on the frequency of the incident field $\omega_{SM}$ and has the width $\Omega_R^2 \gamma_0 |\alpha|^2 / \Delta^2 \ll \gamma_0$. This width is much smaller than the linewidth of the spontaneous emission of the atom. Therefore, at times $\gamma_0^{-1} \ll t \ll \left(\Omega_R^2 \gamma_0 |\alpha|^2 / \Delta^2\right)^{-1}$, we have harmonic oscillations of the dipole moment with the frequency $\omega_{SM}$ of the selected mode with the amplitude $\Omega_R |\alpha| / |\Delta|$ that is exactly equal to the amplitude of the oscillations of the dipole moment of the atom in the problem in which the field is considered classically.

## 6. Why Placzek's theory requires the second order of the perturbation theory?

As pointed above, the theory that considers scattering as two-step process of absorption and emission has internal contradictions and phenomenological assumptions. Many works when conducting quantitative evaluation ignore the presence of a virtual level, but use it to explain the mechanism of the phenomenon under consideration. As this theory is widely used, it is reasonable to consider criticism of the theory in more detail.

Usually, the photon absorption and emission theory of scattering is considered as a special case of the spectroscopic theory of absorption and emission of photons during the transition of an atom between its own states. This theory assumes that the transitions are caused by the incident field. In the first order of the perturbation theory, according to Fermi's golden rule, the probability of the transition between initial and final states, which is caused by the dipole interaction $\hat{V} = -\mathbf{d} \cdot \mathbf{E}$, is proportional to $\left|\langle i|\hat{V}|f\rangle\right|^2$ [8]. Since in the case of elastic scattering, the matrix elements of the dipole moment between the same atomic states, $|f\rangle = |i\rangle$, are zero, $|\langle i|\hat{V}|f\rangle|^2 = 0$, usually one employs the second order of perturbation theory. The resulted formula interprets the elastic scattering of light as of two-steps process. In the first step, the atom being in the ground state absorbs an incident photon and transitions to an excited state, which is a superposition of excited atomic states. In the second step, the excited atom emits a photon.

This approach is based on Fermi's golden rule. This rule suggested that the initial and final states are eigenstates of the system. Thus, the emission-absorption system considers an atom as a system and an incident field as a perturbation.

In reality, as was shown by Janes and Cummings [12], the interaction of an atom with a single mode of an EM field significantly changes the eigenstates of the system. In particular, the characteristic frequency shift between the eigenstates becomes equal to the frequency of the mode.

Since Fermi's golden rule deals with eigenstates of the system, the theory should consider the eigenstates of the "atom + selected mode" system, which are a superposition of atom and field eigenstates [12]. All eigenstates are split into pairs with the fixed number $n$ of photons in the mode, $|+,n\rangle$ and $|-,n\rangle$. If there is only one photon, we can consider only two Jaynes–Cummings eigenstates with $n = 1$. To use Fermi's golden rule, one has to present the initial and final states as a combination of the Janes-Cummings solutions, $|i\rangle = |g,1\rangle = c_1|+,n\rangle + c_2|-,n\rangle$. This presentation corresponds to the complex Rabi oscillations between $|+,n\rangle$ and $|-,n\rangle$ without a stationary state. Indeed, the perturbation that leads to transitions and photon emission is integrated into eigenstates. To obtain the stationary state, $|g,0\rangle$, it is necessary to take into account the interaction of the atom with all free space modes, the fields of which now play the role of a perturbation.

Fermi's rule gives us the probability amplitudes of transitions from $|+,n\rangle$ to $|g,0\rangle$ and $|-,n\rangle$ to $|g,0\rangle$. The probability of the transition from $|i\rangle = |g,1\rangle$ to $|g,0\rangle$ is proportional to the sum $|c_1\langle +,1|g,0\rangle|^2 + |c_2\langle -,1|g,0\rangle|^2$, while Placzek's theory gives $|c_1\langle +,1|g,0\rangle + c_2\langle -,1|g,0\rangle|^2 = |\langle g,1|g,0\rangle|^2$. The former sum is not zero, while the latter one is. Therefore, the scattering process should be considered as a first-order radiation process.

## 7. Conclusion

In this paper, we treat the scattering process as a process of relaxation of the system consisting of an atom and an incident field into free space modes. We model the incident field as a mode selected from the set of free space modes of the EM field and the set of the initial number of photons in this selected mode. The eigenstates of the system are a superposition of the excited states of the atom and the mode. As an initial condition, we consider the state, which is realized in experiment. In this state, the atom is in the ground state, but the selected mode is excited. Since the initial state is not a system eigenstate, the system density matrix oscillates, and the system periodically transitions to its eigenstates. Then, from this superposition state, the system passes to another superposition state, emitting a photon. This process continues until no photons remain in the selected mode. Due to the oscillations of the system between its eigenstates, the consideration of the scattering process as a two-photon process is not adequate because there is no state in which the initial photon is completely absorbed.

We carry out a comparative analysis of the various initial conditions, which are commonly used in the literature. We demonstrate that if only one photon is initially in the selected mode (the Berestetskii-Lifshitz-Pitaevskii approach [6]), then the radiation spectrum has an extremely narrow line. However, the system radiation is extremely slow, and the total radiated energy is also small. Note that this approach describes the experimental situation when only one photon interacts with the atom. It can be achieved if the external beam is extremely attenuated.

In Placzek's approach [8], the initial number of photons is large so that the field at the atom's location is measurable. The analysis of this approach shows that a cascade of transitions between the eigenstates of the "selected mode + atom" system leads to dynamic equilibrium of the atom with a constant population of the excited level. An atom in this state continuously emits photons that form a spectrum centered on the frequency of the selected mode with a linewidth that is much smaller than the width of the line of the spontaneous relaxation of the atom. However, radiated light is incoherent. This approach describes the experimental situation when an external beam is incoherent, e.g., the beam from a mercury-vapor lamp.

To obtain a coherent response, one needs to have a coherent initial state of the mode with a large number of photons. The latter actually means that the field can be considered to be classical. This approach describes a common experimental situation in which coherent laser light is used.

To conclude, we develop a microscopic theory of Rayleigh scattering of light by atoms that describes the main features observed in experiment. This includes the elasticity of the scattering and coherence of the scattered light. The theory does not use any phenomenological assumptions, such as an additional virtual level that has been widely used in previous theoretical approaches.

**Acknowledgments**

A.A.L. acknowledges the support of the ONR under Grant No. N00014-20-1-2198. I.V.D and E.S.A. were supported by the Foundation for the Advancement of Theoretical Physics and Mathematics BASIS.